\documentclass[aps,prd,twocolumn,nofootinbib,notitlepage,supersciptaddress,floatfix]{revtex4-1}
\usepackage[colorlinks=false, linkcolor=black, citecolor=black]{hyperref}
\usepackage{graphicx,color}
\usepackage{subfig}
\usepackage{amstext,amsmath,amssymb,amsfonts,amsthm}
\usepackage[format=plain,justification=centerlast]{caption}

\def\be{\begin{equation}}
\def\ee{\end{equation}}
\def\bes{\begin{eqnarray}}
\def\ees{\end{eqnarray}}
\def\ba{\begin{align}}
\def\ena{\end{align}}
\def\bwt{\begin{widetext}}
\def\ewt{\end{widetext}}

\begin{document}

\title{Backreaction due to quantum tunneling and modification to the black hole evaporation process}
\date{\today}
\author{Sujoy K. Modak}
\email{sujoy.kumar@correo.nucleares.unam.mx}
\affiliation{ Instituto de Ciencias Nucleares, Universidad Nacional Aut\'{o}noma de M\'{e}xico, Apartado Postal 70-543, Distrito Federal, 04510, M\'{e}xico}

\begin{abstract}
We study the effect of backreaction on the evaporation of quantum black holes. The method used is based on quantum tunneling formalism as proposed in \cite{Banerjee:2009wb}. We give a more realistic picture by considering the fact that a black hole looses its energy while modes are tunneled outside the event horizon. It is shown how the tunneling quantum field modes affect the geometry and how this change in geometry is arrested in the quantum field. Exploiting this we calculate the modified (nonthermal) radiation spectrum, associating energy fluxes and discuss various issues related with these. The results obtained here are often expected on physical grounds, but, importantly we find them in a {\it quantitative} manner.
\end{abstract}

\maketitle

\section{Introduction}
Hawking's original approach \cite{H2} of black hole radiation is based on a global set up of collapsing geometry where the asymptotic {\it in} vacuum state is realized as a particle state in the asymptotic {\it out} vacuum. Such an evolution is believed to be non-unitary in the sense that a {\it pure} initial state evolves as a {\it mixed} state in the out region. In the late time limit when the black hole settles down one obtains a thermal (blackbody) radiation spectrum. While this picture is technically rigorous it does not consider the evaporation of black holes by incorporating the backreaction of the particles (on the spacetime and vice-versa) that are being emitted. Therefore one lacks a physical picture where the energy-momentum of the combined system is conserved. In order to do that one should see the effect of particle emission as a loss of mass energy, electrostatic energy as well as rotational energy for the charged-rotating black holes. In a more realistic picture one would expect these black hole parameters are no longer constant, rather, they vary while particles are being emitted. This should naturally modify the background geometry and one can justify the origin of created particles.

However, incorporating the effect of backreaction have not always been easy. It may vary depending on the concerned approach to the problem. For an arbitrary spacetime, the most natural way to consider the backreaction effect is to modify the Einstein equation in its semiclassical form $G_{\mu\nu} = 8\pi G \langle T_{\mu\nu} \rangle$. Where in the right hand side one includes the vacuum expectation value of the energy momentum tensor. In this way one includes the quantum effects of the field to the spacetime geometry. However, more often than not, this equation is difficult to solve analytically. Even for two dimensional case one needs a numerical handle for this \cite{Ashtekar:2010qz}. In fact there is no preferred methodology to consider every aspects of backreaction in practice. At most one can come up with an useful approach in case by case to include some aspects of backreaction. 

For example, in regard to the quantum emission by black holes, the effect of backreaction was partially addressed by Parikh and Wilczek \cite{Parikh:1999mf}, where, in the simplest case, the emission of particle of energy ($\omega$) actually reduces the mass of the black hole by the same amount. This modifies the tunneling probability, which is otherwise thermal, by bringing nonlinear terms of $\omega$ in the expression of tunneling probability. However since this approach is only intended to derive Hawking temperature one fails to obtain the radiation spectrum in presence of backreaction. Noteworthy, radiation spectrum in presence of backreaction is important to go beyond the kinematic approximation of Hawking radiation. Usually, when the energy flux deviates from the thermal nature one finds correlations in the outgoing radiation. These correlations then give us an estimation of information that was hiding behind the horizon. In this way we recover some information about the matter that originally created the black hole. One might also be interested to argue that actually there will be complete recovery of information due to backreaction and there will not be any {\it information loss}. However, a recent work on CGHS black holes \cite{Ashtekar:2010qz} has argued that actually backreaction {\it does not} recover all the information. Even though, it is a fact that correlations in the outgoing radiation allow us to know an {\it incomplete} information about the matter that created the black hole. Therefore backreaction is an important aspect to look on, especially, in the absence of a quantum theory of gravity. 

In this paper we present a simple methodology to treat the backreaction effect on the outgoing radiation in a black hole spacetime. For simplicity we consider Schwarzschild like spacetime where an emission of quanta of energy $\omega$ actually reduces the mass of the black hole by the same amount. Our approach is based on the work \cite{Banerjee:2009wb} where a reformulation of quantum tunneling method was used to derive radiation spectrum for black holes. Later this method was also used in other cases \cite{Banerjee:2009sz}-\cite{Sakalli:2012zy}. However, all these works have neglected backreaction due to created particles. 

There exist some efforts \cite{Jiang:2010ma}-\cite{Corda:2013eua} to include backreaction effect making host the method of \cite{Banerjee:2009wb}. However, as we see, this attempt, although was reasonable, by construction it  has very slight departure from thermality. Therefore the radiation spectrum that was found in \cite{Jiang:2010ma, Jiang:2010ra} is necessarily thermal (with a modified temperature). Nevertheless, such a result is unexpected if backreaction effect is large enough as shown explicitly in \cite{Ashtekar:2010qz} for CGHS black hole. The departure from thermal behavior is thus expected to be a natural consequence of backreaction. 

Here we show that an appropriate modification of \cite{Banerjee:2009wb} also allows to include large backreaction effect (the energy scale, however, is fairly lower than the Planck scale) due to created particles. We discuss the fact that when particles are being emitted by black holes, energy of the latter goes down which is also imprinted on the state of the quantum field. These back-reacted quantum field state, after correct interpretations, provides modifications to the standard thermal (uncorrelated) radiation spectrum. The new spectrum of particle distribution as found here is non-thermal since it includes nonlinear terms in frequencies $\omega$. In practice, it is difficult to get a closed expression for that and in this paper we give an order by order approximation of the desired result within analytic limit. Finally we provide a graphical analysis to show physical essence of backreaction effect as applied to the emission spectrum and radiated energy associated with it.

\section{Review of the basic setup}
We start by reviewing the standard approach based on WKB picture of tunneling formalism as given in \cite{Banerjee:2009wb}. This gives a intuitive understanding of particle emission and leads to the thermal radiation spectrum in a simplest manner. Later we shall modify this picture to accommodate the backreaction effect and find out the modification on the radiation spectrum. 
\subsection{WKB ansatz for field modes}
Let us consider the simplest form of a black hole metric given by the spherically symmetric form
\begin{equation}
ds^2=-f(r)dt^2 + \frac{1}{f(r)}dr^2 + r^2d\Omega^2.
\label{mtrc}
\end{equation}
For simplicity we shall consider Schwarzschild black hole with $f(r)=(1-\frac{2M}{r})$ and study the emission of masslees and chargeless scalar particles. Also since in tunneling method one is concerned with the near horizon geometry it is only the $r,t$ part of the metric that govern the dynamics of the quantum fields. The scalar field equation in the background of (\ref{mtrc}), is given by
\begin{equation}
-\frac{\hbar^2}{\sqrt{-g}}{\partial_\mu[g^{\mu\nu}\sqrt{-g}\partial_{\nu}]\Phi}=0.
\label{scfeq}
\end{equation}
The WKB ansatz for the scalar field has the form
\begin{eqnarray} 
\Phi(r,t)=\exp[-\frac{i}{\hbar}{{\cal S}(r.t)}], 
\label{scfan}
\end{eqnarray}
where the action is expanded in powers of $\hbar$
\begin{eqnarray}
{\cal S}(r,t)={\cal S}_{0}(r,t)+\displaystyle\sum_{i=1}^{\infty}\hbar^{i}{\cal S}_{i}(r,t).
\label{sfaction}
\end{eqnarray} 
Now substituting (\ref{sfaction}) into (\ref{scfan}) we get
\begin{eqnarray}
&&\frac{i}{{f(r)}}\Big(\frac{\partial {\cal S}}{\partial t}\Big)^2 - if(r)\Big(\frac{\partial {\cal S}}{\partial r}\Big)^2 - \frac{\hbar}{f(r)}\frac{\partial^2 {\cal S}}{\partial t^2} + \hbar f(r)\frac{\partial^2 {\cal S}}{\partial r^2}
\nonumber
\\
&&+ \hbar{\frac{\partial f(r)}{\partial r}}\frac{\partial {\cal S}}{\partial r}=0.
\label{2.3}
\end{eqnarray}
Taking the semi-classical limit ($\hbar\rightarrow 0$), we obtain the first order partial differential equation,
\begin{eqnarray}
\frac{\partial {\cal S}_0}{\partial t}=\pm {f(r)}\frac{\partial {\cal S}_0}{\partial r}.
\label{semic}
\end{eqnarray}
This is identified with the semi-classical Hamilton-Jacobi equation. Therefore the ansatz for the semi-classical action for a scalar field moving under the background metric (\ref{mtrc}) should be chosen following semi-classical Hamilton-Jacobi theory. For that one refers to the time translational symmetry of (\ref{mtrc}) writes the action in the form
\begin{eqnarray}
{\cal S}_{0}(r,t)=\omega t+{\cal S}_{0}(r),
\label{actionformm}
\end{eqnarray} 
where $\omega$ is the energy of the quantum field. Substituting this in (\ref{semic}) it yields 
\begin{eqnarray}
{\cal S}_{0}(r)=\pm\omega\int\frac{dr}{f(r)},
\label{s0n}
\end{eqnarray}
Using (\ref{s0n}) in (\ref{actionformm}) the semiclassical action is found to be
\begin{eqnarray}
{\cal S}_{0}(r,t)=\omega(t\pm\int\frac{dr}{f(r)}).
\label{action2}
\end{eqnarray} 
Therefore the solution for the scalar field (\ref{scfan}) is given by 
\begin{eqnarray} 
\Phi(r,t) &=& e^{[-{\frac{i}{\hbar}}\omega(t\pm\int\frac{dr}{f(r)})]}\nonumber\\
          &=& e^{-\frac{i}{\hbar}\omega(t\pm r^*)} 
\label{soln}
\end{eqnarray}
where the tortoise coordinate is defined as $r^*=\int\frac{dr}{f(r)}$. The $+ (-)$ sign stands for the right (left) moving modes. While left moving modes reaches singularity the right moving tunnels outside the horizon. 

\subsection{Relation between {\it in}, {\it out} modes}
The outgoing mode while tunnels out the horizon from inside to outside the definition of coordinate changes in Schwarzschild system. As explained in \cite{Banerjee:2009wb} using Kruskal coordinates one can set up a relation between the inside and outside Schwarzschild patches which then express the inside mode with respect to an observer outside the event horizon. 

Kruskal time and space coordinates inside and outside the horizon are defined as
\bwt
\begin{eqnarray}
T_{in} = \exp(\kappa r^{*}_{in}) \cosh(\kappa t_{in});~ X_{in}=\exp(\kappa r^{*}_{in})\sinh(\kappa t_{in}) \\
T_{out} = \exp(\kappa r^{*}_{out})\sinh(\kappa t_{out});~ X_{out}= \exp(\kappa r^{*}_{out}) \cosh(\kappa t_{out}).
\label{capital4}
\end{eqnarray}
\ewt
These two sets of coordinates can be swapped with each other by the following transformations
{\begin{eqnarray}
t_{in}\rightarrow t_{out}-\frac{i\pi}{2\kappa}\nonumber\\
r^*_{in}\rightarrow r^*_{out}+\frac{i\pi}{2\kappa}.
\label{connect1}
\end{eqnarray}
With these identifications the scalar field modes are easily transformed as 
\begin{eqnarray}
\Phi_{in}^{(R)} \rightarrow {e^{-\frac{\pi\omega}{\hbar\kappa}}}\Phi_{{out}}^{{(R)}} \nonumber\\
\Phi_{in}^{(L)} \rightarrow {\Phi_{{out}}^{{(L)}}}
\label{connect3}
\end{eqnarray} 

\subsection{Density matrix and radiation spectrum}
A particle state corresponding to an arbitrary but finite number of virtual pairs inside the black hole event horizon,
\begin{eqnarray}
|\psi\rangle = N \displaystyle\sum_{n} |n^{(L)}_{\textrm{in}}\rangle\otimes|n^{(R)}_{\textrm{in}}\rangle
\label{pstatein}
\end{eqnarray}
where $N$ is a normalization constant and $\kappa (M)=\frac{1}{4\pi M}$ is the surface gravity at horizon.  This state has finite numbers of particles with particular frequencies but not {\it all} of the frequencies have non-zero particles, otherwise, the sum $\sum n \hbar \omega$ will diverge making $|\psi\rangle$ a non-physical state. While transforming this quantum state with respect to an observer outside the horizon this takes the form
\begin{eqnarray}
|\psi\rangle = N \displaystyle\sum_{n} e^{-\frac{\pi n\omega}{\hbar\kappa(M)}}|n^{(L)}_{\textrm{out}}\rangle\otimes|n^{(R)}_{\textrm{out}}\rangle.
\label{pstateout}
\end{eqnarray}
The normalization constant is found from $\langle \psi | \psi \rangle=1$, given by
\begin{eqnarray}
N^2 = \frac{e^{\beta  \omega }}{e^{\beta  \omega - 1}}
\label{N0}
\end{eqnarray}
To obtain the average particle number in the energy state $\omega$ with respect to an outside observer, one defines the density matrix operator $\rho= |\psi\rangle\langle\psi|$ and then traces out the inside degrees of freedom to yield the reduced density matrix of the form
\begin{eqnarray}
\rho_{red}=\left(1-\exp(-\frac{2\pi\omega}{\hbar\kappa})\right) \displaystyle\sum_{n=0}^{\infty} e^{-\beta n\omega} |n^{(R)}_{\textrm{out}}\rangle\otimes\langle n^{(R)}_{\textrm{out}}|.\nonumber\\
\end{eqnarray}
Consequently the number distribution with respect to $\omega$ is given by $\langle n_{\omega} \rangle = trace(n\rho_{red})$. This is found to be Planckian
\begin{eqnarray}
\langle n_{\omega} \rangle = \frac{1}{-1+e^{\beta  \omega }}
\label{n0}
\end{eqnarray}
with the temperature of radiation given by the Hawking temperature $T=\frac{1}{\beta}=\frac{\hbar\kappa}{2\pi}$. The energy flux reaching the asymptotic observer then matches with the {\it uncorrelated} Hawking flux.


\section{Radiation Spectrum with Backreaction}

Although the last section gives a reasonable methodology to derive the radiation spectrum, it does not take into account the backreaction of the matter field on the black hole spacetime. In other words the energy-momentum conservation is not satisfied. Now we shall modify the standard methodology by considering the fact that mass of the Schwarzschild black hole should be reduced to an amount equal to the energy carried out by emitted radiation.

\subsection{Effect of backreaction}
In order to include backreaction we consider the fact, that, black hole's mass $M$ is reduced to $M-n \hbar \omega$ once $n$ outgoing modes with energy $\hbar \omega$ are tunneled outside the horizon. Equivalently, an identical number of negative energy modes are excited inside the horizon and reduces the hole's mass. As a result the surface gravity changes from its earlier value when the black hole was not emitting. Given this argument one can realize a direct modification to the relations between the inside and outside coordinate patches in (\ref{connect1}). For Schwarzschild type black holes this is implemented by redefining the surface gravity $\kappa(M)=\frac{1}{4 M} \rightarrow \kappa(M-n\hbar \omega)= \frac{1}{4(M-n\hbar \omega)}$ in (\ref{connect1}). This in fact naturally transfers the effect of backreaction in the transformation of modes as given in (\ref{connect3}). The only thing we need to do is to use the new expression of the surface gravity with effective mass of the hole. Therefore, the quantum state (\ref{pstateout}) is modified in the following way
\begin{eqnarray}
|\psi\rangle = N \displaystyle\sum_{n=0}^{\infty} e^{-\frac{4\pi n \omega(M-n\hbar \omega)}{\hbar}}|n^{(L)}_{\textrm{out}}\rangle\otimes|n^{(R)}_{\textrm{out}}\rangle.
\label{ms}
\end{eqnarray}
Before going further let us highlight some features of the above expression for the quantum state. Due to the new term in the parenthesis we have effectively a sum over positive quadratic power ($n^2\omega^2$) on exponential which clearly diverges once we sum over $n$ in the limit $0 \le n \le \infty$. At this point one might be worried as to what extent oneself should trust the above definition. To get an answer one should realize that such a behavior is not unexpected given the fact that during the late stage of Hawking radiation, when mass of the black hole approaches zero, the process of evaporation becomes more violent releasing an enormous amount of energy with an infinite rate. Such a process off-course not expected in physical ground which again teaches us that at certain level the process of Hawking effect should be over-taken by a Quantum Gravity (QG) theory. It is exactly the situation with the modified quantum state (MQS) given in (\ref{ms}), which, expectantly giving us a hint that we should not trust the analysis based on this in the QG region. The question is what should be the scale up to which this quantum state provides us an acceptable physical picture? We propose that this analysis should make sense until the mass of the black hole is well above the Planck mass, say, for example, a thousand time the Planck mass. Then the total energy carried away by Hawking evaporation is $\displaystyle\sum_{n=0}^{n_{max}} n \hbar \omega = M- 1000 m_p$. It is now clear that the upper limit $n_{max}$ is directly related with the original mass of the black hole ($M$). For an astrophysical black hole $M$ being very large one can take $n_{max} \approx \infty$ and still the MQS (\ref{ms}) will work. However, from comparatively smaller black holes it may not be the case and one should be careful enough to use the MQS. More precisely, either we should find out a way to put a number for $n_{max}$ by some rigorous analysis so that we treat the near QG region correctly, or, alternatively, find out an approximate analysis with $n_{max} = \infty$ which is only intended for the energy scale much greater than Planck energy. In this paper we shall follow the second route which is easier to tackle, nevertheless, provide important results and understanding on backreaction effect in a quantitative manner. We have more to say on this using numerics in the next section.

In order to attack the problem analytically, we do not put any cut off on $n$, rather series expand the quadratic positive power in the exponential in (\ref{ms}) so that
\begin{eqnarray}
|\psi\rangle = N \displaystyle\sum_{n=0}^{\infty} e^{-\frac{4\pi n \omega}{\hbar}} \left(1+\displaystyle\sum_{l=1}^{\infty}\frac{(4\pi \omega^2 n^2)^{l}}{l!}\right) |n^{(L)}_{\textrm{out}}\rangle\otimes|n^{(R)}_{\textrm{out}}\rangle. \nonumber\\
\label{mss}
\end{eqnarray}
The advantage of the above expression is that the term which causes the divergence can now be treated in an order by order expansion in $l$. As we mentioned, the effect of this series expansion, on the radiation spectrum or energy flux crucially depends on the original mass of the black hole. For a larger mass black hole ({\it e.g.} astrophysical), the effect of backreaction on Hawking radiation, in the lowest orders in $l$ should be much weaker than a comparatively lesser mass ({\it e.g.} microscopic) black hole. We shall discuss these aspects shortly in the next section.

In order to proceed further we first calculate the reduced density matrix using (\ref{ms}) and then expand it in the following manner
\begin{eqnarray}
\rho_{red} &=& N^2 \displaystyle\sum_{n=0}^{\infty} e^{-\frac{8\pi n \omega(M-n\hbar \omega)}{\hbar}} |n^{(R)}_{\textrm{out}}\rangle\otimes\langle n^{(R)}_{\textrm{out}}| \label{dmf}\\
                 &=& N^2 \displaystyle\sum_{n=0}^{\infty} e^{-\frac{8\pi n \omega M}{\hbar}}  e^{8\pi \omega^2 n^2} |n^{(R)}_{\textrm{out}}\rangle\otimes\langle n^{(R)}_{\textrm{out}}| \nonumber\\
                &=& N^2  \displaystyle\sum_{n=0}^{\infty} e^{-\beta n\omega}\left(1+\displaystyle\sum_{l=1}^{\infty}\frac{(8\pi \omega^2 n^2)^{l}}{l!}\right)  |n^{(R)}_{\textrm{out}}\rangle\otimes\langle n^{(R)}_{\textrm{out}}| \nonumber \\
\label{dm}
\end{eqnarray}
 where the inverse temperature $\beta = \frac{8\pi M}{\hbar}$. It is now possible to find the modified radiation spectrum by an order by order expansion over $l$. From the above discussion it is now clear that the approximation $M >> n \hbar\omega$ is equivalent to ignoring backreaction of the quantum field on the spacetime. We shall avoid this assumption by considering higher order terms in $l$. However, by restricting to a few terms in the series of $l$ will eventually imply that the backreaction effect is comparatively weaker and valid only in the initial stages of evaporation such that energy scale is fairly lower than the QG regime.

\subsection{Analytical results for radiation spectrum in order by order expansion}
In principle, to consider the effect of backreaction one should use the state (\ref{ms}) or equivalently the density matrix (\ref{dmf}). However, given the restrictions discussed in the last sub-section, in this work we use (\ref{dm}) and find the desired expressions for the radiation spectrum by an order by order expansion for cases $l=1$ to $l=3$. As mentioned above these expressions are therefore valid only in the initial stages of black hole evaporation. The results of this subsection will be used in the next section to discuss main physical implications.

We prefer not to repeat the analysis which is basically similar that presented in the last section. Only the final results for various expressions are provided. In the subsequent expressions we set $\alpha = 8\pi \omega^2$ that will appear in many places and this is a signature of the backreaction effect.

(i) {\it{$l= 1$ (up-to leading order correction to the radiation spectrum):}} The normalization constant is found  from $\langle\psi | \psi\rangle =1$, given by 
\begin{eqnarray}
N^2 = \frac{e^{\beta  \omega } \left(\left(-1+e^{\beta  \omega }\right)^2 +\alpha \left(1+e^{\beta  \omega }\right)\right)}{\left(-1+e^{\beta  \omega }\right)^3},
\label{N1}
\end{eqnarray}
where the quantum state $|\psi\rangle$ is now defined in (\ref{ms}). The reduced density matrix as given in (\ref{dm}) is now nonthermal. The final expression for radiation spectrum is given by
\begin{eqnarray}
\langle n_{\omega} \rangle = \frac{2e^{\beta\omega}[-1+\cosh(\beta\omega)+\alpha(2+\cosh(\beta\omega))]}{(e^{\beta\omega}-1)[(e^{\beta\omega}-1)^2+\alpha(e^{\beta\omega}+1)]}
\label{n1}
\end{eqnarray}
It is reassuring to notice that in the limit $\alpha \rightarrow 0$ we get back (\ref{n0}) as expected for the non-backreacting case.

(ii) {\it{$l= 1, 2$ (up-to second order correction to the radiation spectrum)}}: Normalization constant,
\begin{eqnarray}
N^2 &=& \frac{e^{\beta\omega}}{\left(e^{\beta \omega}-1\right)}\left(1+\alpha \frac{e^{\beta\omega} +1}{\left(e^{\beta  \omega }-1\right)^2}\right. \nonumber\\
&&\left. +\alpha^2 \frac{(1+e^{\beta\omega})(1+10e^{\beta\omega} + e^{2\beta\omega})}{(e^{\beta\omega}-1)^4}  \right) \label{N2}
\end{eqnarray}
Average particle number in frequency $\omega$,
\bwt
\begin{eqnarray}
\langle n_{\omega} \rangle = \frac{e^{2\beta\omega}[(6-6\alpha + 33\alpha^2) + (-8+4\alpha + 26\alpha^2)\cosh(\beta\omega)+(2+2\alpha + \alpha^2)\cosh(2\beta\omega)]}{(-1+e^{\beta\omega})^3[(e^{\beta\omega}-1)^2+\alpha(e^{\beta\omega}+1)]+ 2 \alpha^2 e^{\beta\omega} \sinh[\beta\omega](5+\cosh(\beta\omega))} 
\label{n2}
\end{eqnarray}
\ewt

(iii) {\it{$l= 1, 2,3$ (up-to third order correction to the radiation spectrum)}}: Normalization constant,
\begin{eqnarray}
N^2 &=& \frac{e^{\beta \omega}}{6x^7}\left(6 x^6 + 6 \alpha x^4 (1+ e^{\beta\omega}) \right. \nonumber\\
&&\left. + 3 \alpha^2 x^2 h_1(\beta, \omega) (1+ e^{\beta\omega}) + 4\alpha^3 h_2(\beta, \omega)  \right)
\end{eqnarray}
Average particle number in frequency $\omega$,
\begin{eqnarray}
\langle n_{\omega} \rangle = \frac{6 x^6 + 3 \alpha x^4 f_1 (\beta,\omega) + \alpha^2 x^2 f_2 (\beta,\omega) + \alpha^3 f_3 (\beta,\omega)}{6 x^7 + 6 \alpha x^5 (1 + e^{\beta \omega}) +3 \alpha^2 x^3 f_4(\beta,\omega) + \alpha^3 x f_5 (\beta,\omega)}\nonumber\\
\label{n3}
\end{eqnarray}

\noindent In the above expressions various functions $x$, $h_1, h_2$ and $f_1$ to $f_{5}$ are defined in Appendix \ref{sec:A}. Note that for all cases, by neglecting nonlinear terms in $\omega$ (equivalently $\alpha$), one gets back the black-body radiation spectrum which is reassuring. 

\section{Backreaction and physical observables}
Now let us use the results obtained so far to discuss physical aspects of the backreaction effect.  This are enlisted below:

\begin{enumerate}
\item{{\it Modified spectrum:} A direct consequence of backreaction is a modification to the thermal (black-body) spectrum due to higher order terms in $\omega$ appearing in the expressions of modified number distributions (\ref{n1}), (\ref{n2}) and (\ref{n3}). These expressions allow us to extend the semi-classical treatment beyond standard kinematic approximation of Hawking radiation. In other words, in the beginning of Hawking radiation while the mass of the black hole is very large the energy carried away by the matter fields is negligible and well-known black-body nature emerges. As the mass shrinks, energy carried out by emitted particles, are no longer negligible and it affects the mass which in turn changes the black-body nature. To get an insight we refer the reader to consult Fig. \ref{backreacted-n}. The non-backreacting case is $l=0$ and others include backreaction {\footnote{Note that there is an infra-red divergence in $\langle n_{\omega} \rangle$ for all which is usually discarded on physical ground.}} The change in $\langle n_{\omega} \rangle$, for lowermost values of $l$, is negligible for black hole of astrophysical size. To show this in Fig. \ref{backreacted-n} we plot  $\langle n_{\omega} \rangle$ for two smaller mass black holes. One with mass $M = 1$ and radius $r_h = 2$ (in natural units){\footnote{In S.I units $M \approx 10^8 M_p$ and radius $r_h = \frac{2G}{c^2} = 1.48 \times 10^{-27}~m \approx 10^8 L_{p}$ since Planck length $L_p = \sqrt{\frac{\hbar G}{c^3}} = 1.616 \times 10^{-35} m$ and Planck mass $M_p = \sqrt{\frac{\hbar c}{G}} =2.176 \times 10^{-8} kg$.}} and the other, a hundred times smaller (with mass $M= 0.01$).  Although both have microscopic size the relatively larger mass black hole deviates a little in its particle content due to backreaction effect at the initial stage considered in this work. However, the other black hole of mass $M = 0.01$ shows significant departure from non-back-reacting particle emission spectrum. This black hole, although $10^6$ times heavier than Planck mass, emits relatively high energy particles as a result of backreaction, even in the lowest order of backreaction considered here (with $1 \le l \le 3$). The rate of emission is therefore diverging with the mass closing further near the Planck region. This is somewhat expected from the semi-classical picture with backreaction. As we mentioned for large value of $l$ one is going to encounter a situation, precisely where, the semi-classical picture breaks-down giving its way to QG. Finally, it is reassuring to notice that asymptotically $\langle n_{\omega} \rangle$ reaches zero for all cases making the particle count for the MQS (\ref{ms}) finite for finite values of $l$.}

\begin{figure*}[t]
\includegraphics[scale=0.6]{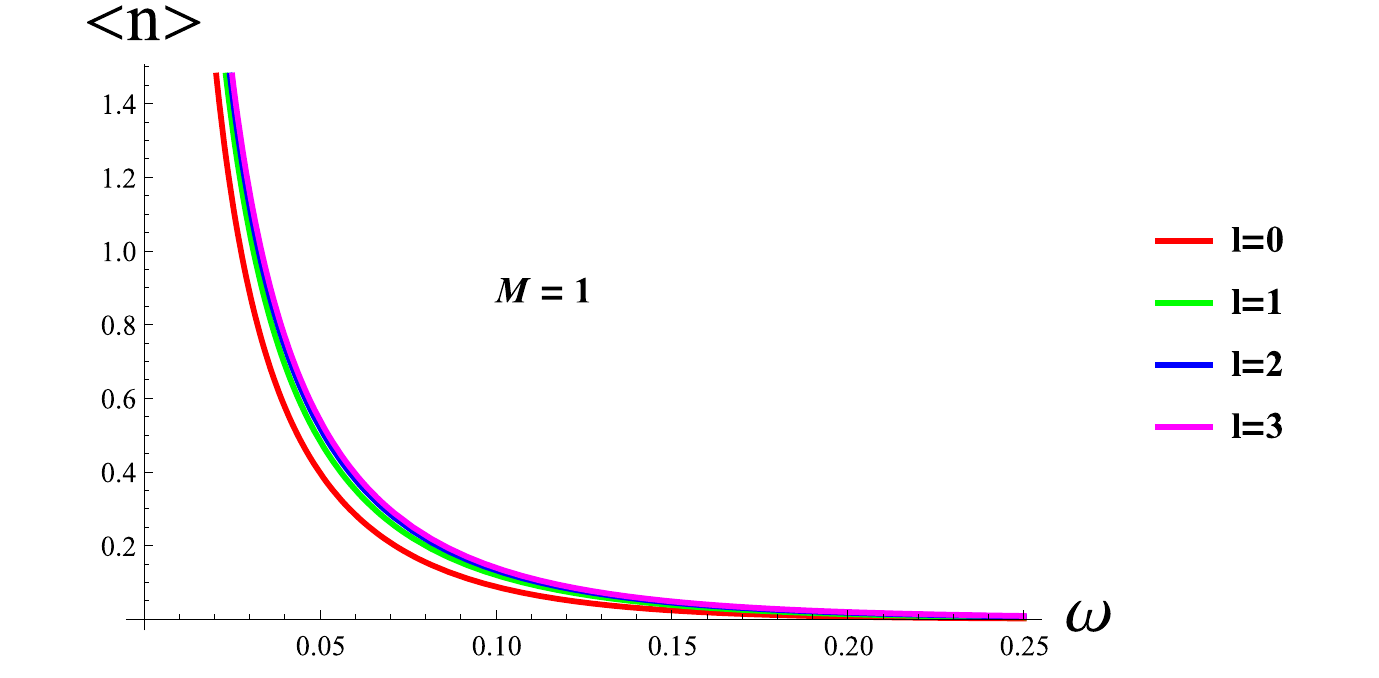}
\includegraphics[scale=0.6]{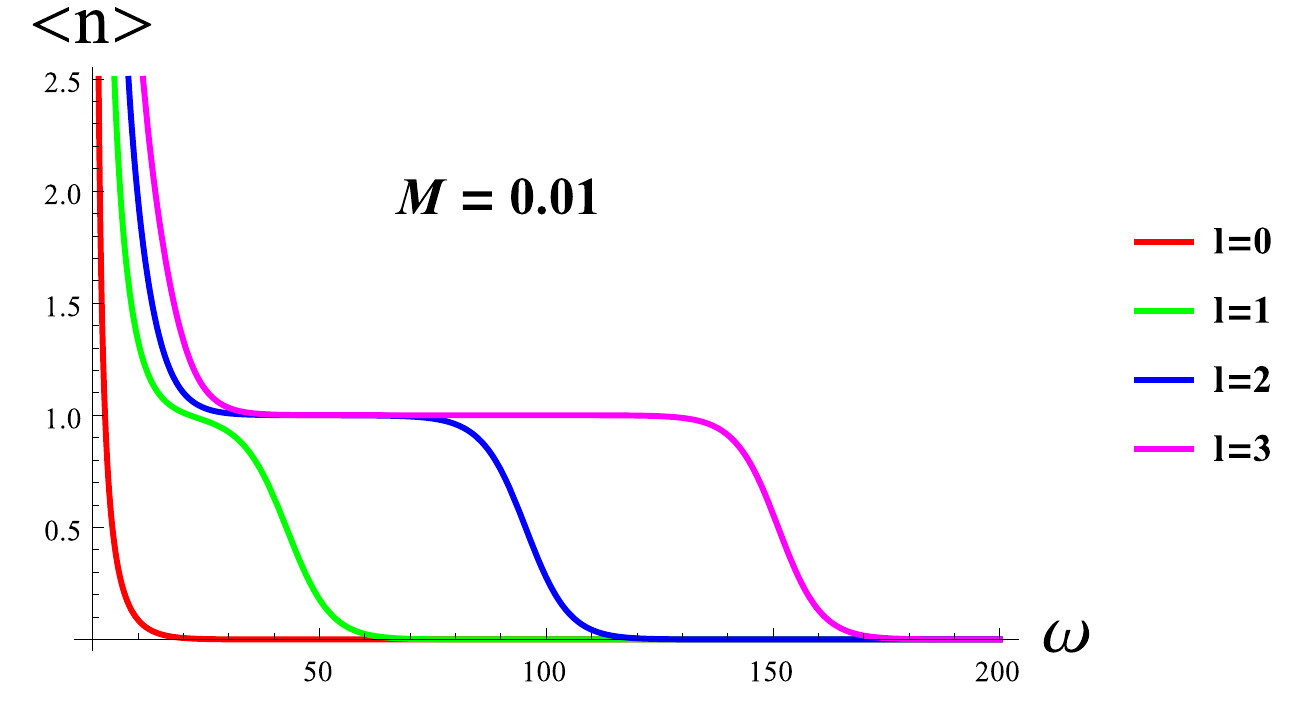}
\caption{Average number of particles $\langle n_{\omega} \rangle$ with respect to frequency $\omega$ without ($l=0$) and with backreaction ($l\ne 0$) effect. In the first plot we assumed $M =1$ while for the other $M=0.01$. These values of $M$ correspond to temperatures $\frac{1}{\beta} = 1.227\times 10^{23} K$ and $\frac{1}{\beta} = 1.227\times 10^{25} K$ respectively. For a detailed discussion see the text.}
\label{backreacted-n}
\end{figure*}

\item{{\it Outgoing correlation and information recovery:} Clearly the higher order terms in (\ref{n1}), (\ref{n2}) and (\ref{n3})  represent  the existing correlation among the outgoing modes which are eventually detected by an asymptotic observer. These correlations then provide {\it some} information about the matter that created the black hole. However, whether or not {\it all} the information could be recovered is beyond the scope of this paper as it includes an interpretation of what is expected from the spacetime singularity.}

\item{{\it Outgoing energy flux:} } The standard definition of energy flux using the average particle number is given by
\begin{eqnarray}
\langle E \rangle = \frac{1}{2\pi} \int_{0}^{\infty} \langle n_{\omega} \rangle \omega d\omega.
 \end{eqnarray} 
In the absence of backreaction the number distribution is given by \eqref{n0}. Then using the above definition one obtains the standard Hawking flux given by
\begin{equation}
\langle E_{0} \rangle = \frac{\pi}{12 \beta^2} = \frac{1}{768 \pi M^2}.
\label{hf}
\end{equation}
Similarly, we can also calculate the energy flux in presence of backreaction as found from the modified distributions \eqref{n1}, \eqref{n2}, \eqref{n3}. However, it is not possible to get an analytic result with the modified expressions. Therefore we integrate them numerically and plot in Fig. \ref{back-en} together with \eqref{hf}. Again we find that for the comparatively larger black hole of mass $M=1$ the increase in radiated energy is much smaller than the smaller mass black hole with $M = 0.01$. For the latter backreaction drains away much more energy even in the lowest order effect of backreaction. However, here for both cases released energy is much smaller than total energy of the black hole $E_{bh} = M c^2$, for the values of $l$ considered in this paper. As we discussed, one should consider this semi-classical analysis, as long as the energy of the black hole remains much higher than Planck energy during the entire stage of evaporation process.  

\end{enumerate}

\begin{figure*}
\includegraphics[scale=0.6]{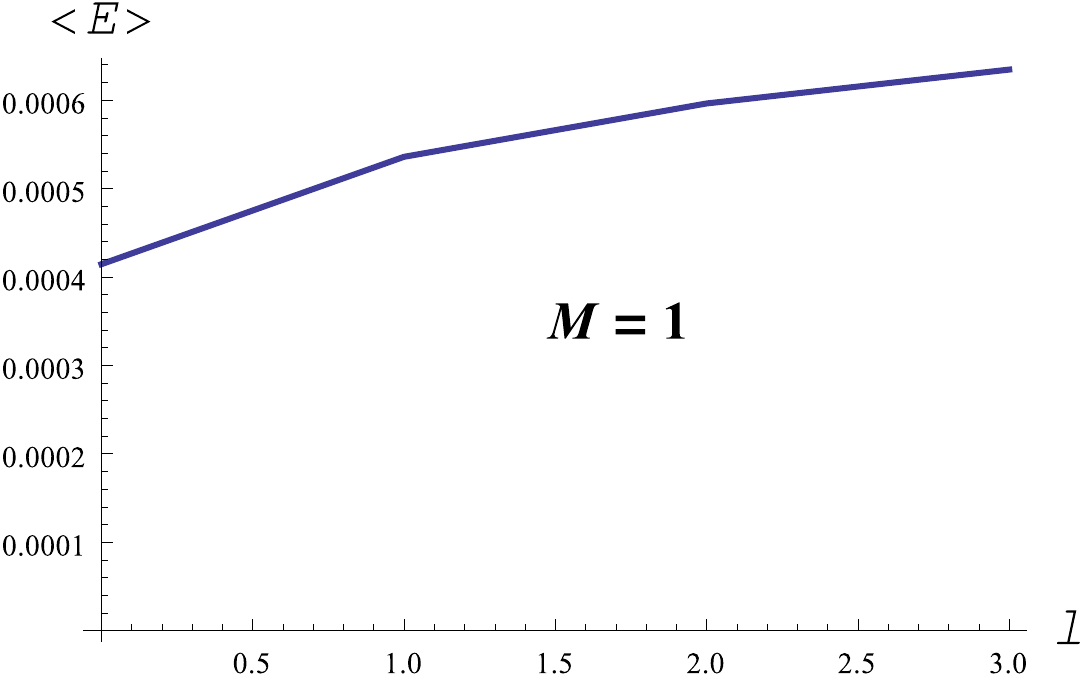}
\includegraphics[scale=0.6]{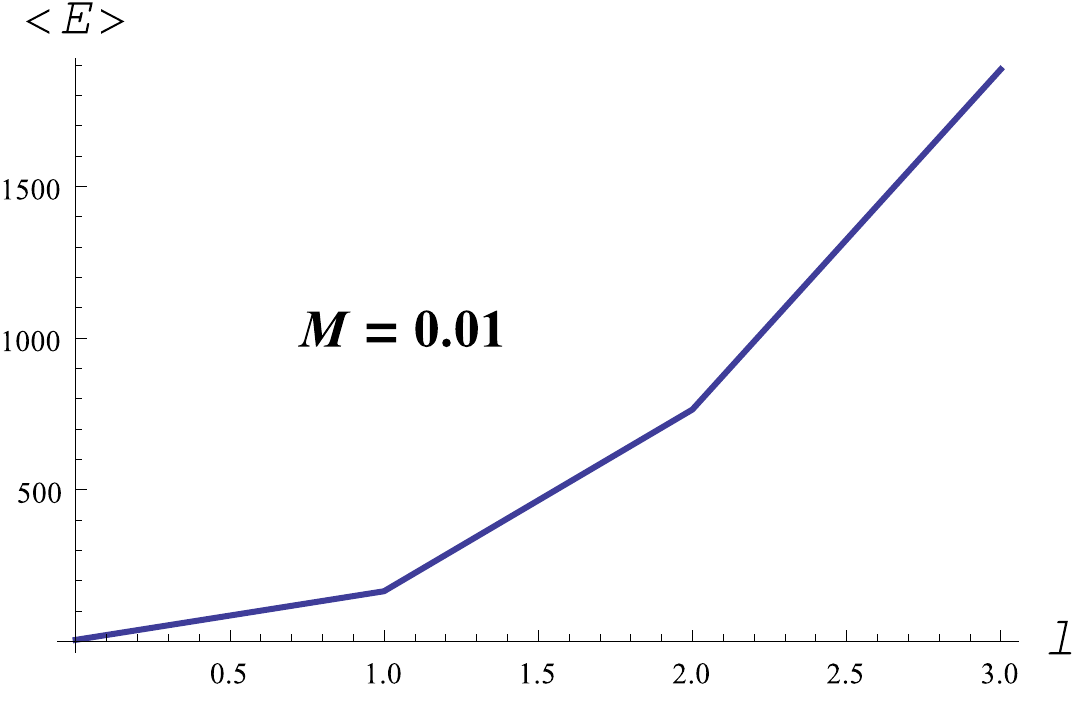}
\caption{Radiated energy flux $\langle E \rangle$ for various $l$. Here $l=0$ is the standard Hawking flux while others with $l\ne 0$ include the backreaction effect. Again these black holes have initial temperatures $\frac{1}{\beta} = 1.227\times 10^{23} K$ and $\frac{1}{\beta} = 1.227\times 10^{25} K$ respectively.}
\label{back-en}
\end{figure*}


\section{Conclusions}
Effect of backreaction on Hawking radiation is a well known but difficult issue to incorporate. There is no well known method which can address this problem to satisfaction even for a simplest black hole geometry in four space-time dimensions. The main problem of using standard picture of Hawking is that it is a global set up which in turn demands a complete knowledge of dynamical evolution. Here we tried to fill this gap by using a reformulation of tunneling method and assuming the natural interpretation of backreaction effect on the state of the quantum field. The advantage of using tunneling method, as used in this paper, is that it is a local phenomenon only sensitive to near horizon geometry. It made our task much simplified. The natural variation in horizon radius/surface gravity was captured in the quantum state. This knowledge was sufficient to find the particle content of the quantum state. 

As anticipated, this modified radiation spectrum included nonlinear terms in frequency $\omega$, which is a signature of correlations in the radiation spectrum. This correlation measuring departure from thermality provides {\it quantum} information which was hiding behind the horizon {\it classically}. We provided an order by order expression in $\omega^2$ for the modified radiation spectrum. We restricted ourselves up to third order, however, there was no problem in computations higher order expressions using our formalism as long as we are far from QG region.

It was found that due to backreaction the black hole emits high energy particles in the late stage of evaporation. A quantitative analysis showed that in the lowest order of backreaction the change in number distribution and energy flux is significant only for microscopic black holes (mass much higher than Planck mass). Smaller black holes or equivalently larger black holes in the late stage of the evaporation process gives off much higher energy particles in presence of backreaction making a significant shift from the standard Hawking flux. Finally, we emphasize that, although, in this paper we have only considered Schwarzschild type black holes, this method can be generalized to charged and rotating cases, by varying other parameters just like the varying mass $M$ here.

\section*{ Acknowledgment}
The author acknowledges useful discussions with Debraj Roy and Douglas Singleton, and encouragement from Sarira Sahu. He also thanks the anonymous referee for providing useful feedback. This work is supported by a DGAPA Postdoctoral Grant from UNAM, Mexico.

\appendix

\section{Complete expressions of various functions ($x$, $h_i$'s and $f_{i}$'s)}\label{sec:A}
Exact expressions for various functions defined in Section III(B) are the following:
\bwt
\begin{eqnarray}
x &=& e^{\beta\omega}-1 \nonumber\\
h_1 &=& 1+ 10e^{\beta\omega} + e^{2\beta\omega}\nonumber\\
h_2 &=& e^{5\beta\omega/2}\cosh[{\beta\omega}/{2}] \left(123 + 56\cosh[\beta\omega] + \cosh[2\beta\omega]\right) \nonumber\\
f_1 &=& 1+ 4 e^{\beta\omega} + e^{2\beta\omega} \nonumber\\
f_2 &=&  8e^{3\beta\omega}\sinh[\beta\omega/2]^2 (33 + 26 \cosh[\beta\omega] + \cosh[2\beta\omega])  \nonumber\\
f_3 &=&  2 e^{3\beta\omega}(1208 + 1191\cosh[\beta\omega] + 120\cosh[2\beta\omega] + \cosh[3\beta\omega])  \nonumber\\
f_4 &=&  (1+e^{\beta\omega}) (1+10e^{\beta\omega}+e^{2\beta\omega})\nonumber\\
f_5 &=& 4 h_2(\beta,\omega) \nonumber
\end{eqnarray}
\ewt

\bibliographystyle{utcaps}

\begin{thebibliography}{10}

\bibitem{H2} S.W.Hawking, ``	 Particle Creation by Black Holes,'' Commun. Math. Phys. {\bf{43}} 199 (1975).


\bibitem{Ashtekar:2010qz} 
  A.~Ashtekar, F.~Pretorius and F.~M.~Ramazanoglu,
  ``Evaporation of 2-Dimensional Black Holes,''
  Phys.\ Rev.\ D {\bf 83}, 044040 (2011)
  [arXiv:1012.0077 [gr-qc]].

\bibitem{Parikh:1999mf} 
  M.~K.~Parikh and F.~Wilczek,
  ``Hawking radiation as tunneling,''
  Phys.\ Rev.\ Lett.\  {\bf 85}, 5042 (2000)
  [hep-th/9907001].

\bibitem{Banerjee:2009wb} 
  R.~Banerjee and B.~R.~Majhi,
  ``Hawking black body spectrum from tunneling mechanism,''
  Phys.\ Lett.\ B {\bf 675}, 243 (2009)
  [arXiv:0903.0250 [hep-th]].
  
  
  \bibitem{Banerjee:2009sz} 
  R.~Banerjee and S.~K.~Modak,
  ``Quantum Tunneling, Blackbody Spectrum and Non-Logarithmic Entropy Correction for Lovelock Black Holes,''
  JHEP {\bf 0911}, 073 (2009)
  [arXiv:0908.2346 [hep-th]].
 
  \bibitem{Roy:2009vy} 
  D.~Roy,
  ``The Unruh thermal spectrum through scalar and fermion tunneling,''
  Phys.\ Lett.\ B {\bf 681}, 185 (2009)
  [arXiv:0908.3149 [hep-th]].
  
  \bibitem{Singleton:2010gz} 
  D.~Singleton, E.~C.~Vagenas, T.~Zhu and J.~-R.~Ren,
  ``Insights and possible resolution to the information loss paradox via the tunneling picture,''
  JHEP {\bf 1008}, 089 (2010)
  [Erratum-ibid.\  {\bf 1101}, 021 (2011)]
  [arXiv:1005.3778 [gr-qc]].
 
 \bibitem{Banerjee:2010rx} 
  R.~Banerjee,
  ``From black holes to emergent gravity,''
  Int.\ J.\ Mod.\ Phys.\ D {\bf 19}, 2365 (2010)
  [arXiv:1005.4286 [gr-qc]].
 
\bibitem{Ren:2010zzc} 
  J.~-R.~Ren, P.~-J.~Mao, R.~Li, T.~Zhu and L.~-Y.~Jia,
   ``Hawking black body spectrum of Goedel black hole from tunneling mechanism,''
  Mod.\ Phys.\ Lett.\ A {\bf 25}, 2167 (2010).
 
 \bibitem{Miao:2011dy} 
  Y.~-G.~Miao, Z.~Xue and S.~-J.~Zhang,
  ``Quantum tunneling and spectroscopy of noncommutative inspired Kerr black hole,''
  Int.\ J.\ Mod.\ Phys.\ D {\bf 21}, 1250018 (2012)
  [arXiv:1102.0074 [hep-th]].
  
 \bibitem{Lorente-Espin:2013kha} 
  O.~Lorente-Espín,
   ``Emission of fermions in little string theory,''
  Phys.\ Rev.\ D {\bf 87}, no. 6, 064016 (2013)
  [arXiv:1303.7098 [hep-th]]. 
  
	\bibitem{Sakalli:2012zy} 
  I.~Sakalli, M.~Halilsoy and H.~Pasaoglu,
  ``Fading Hawking Radiation,''
  Astrophys.\ Space Sci.\  {\bf 340}, 155 (2012)
  [arXiv:1202.3259 [gr-qc]].
     
\bibitem{Jiang:2010ma} 
  Q.~-Q.~Jiang and X.~Cai,
  ``Back reaction, covariant anomaly and effective action,''
  JHEP {\bf 1005}, 012 (2010)
  [arXiv:1005.0999 [gr-qc]].  

\bibitem{Jiang:2010ra} 
  Q.~-Q.~Jiang and X.~Cai,
  ``Back reaction, emission spectrum and entropy spectroscopy,''
  JHEP {\bf 1011}, 066 (2010)
  [Erratum-ibid.\  {\bf 1206}, 118 (2012)]
  [arXiv:1011.0243 [hep-th]].
	
\bibitem{Corda:2013eua} 
  C.~Corda,
   ``Non-strictly black body spectrum from the tunnelling mechanism,''
  Ann.\  Phys.\  337, {\bf 49} (2013)
  [arXiv:1305.4529 [gr-qc]].	

\end{thebibliography}
\providecommand{\href}[2]{#2}\begingroup\raggedright\endgroup
\end{document}